\documentstyle[12pt,twoside]{article}
\newcommand{\br} {{\bf r}}
\newcommand{\bx} {{\bf x}}
\newcommand{\bp} {{\bf p}}
\newcommand{\by} {{\bf y}}
\newcommand{\bz} {{\bf z}}
\newcommand{\bsigma} {{\mbox{\boldmath $\sigma$}}}
\newcommand{\balpha} {{\mbox{\boldmath $\alpha$}}}
\newcommand{\bnabla} {{\mbox{\boldmath $\nabla$}}}
\newcommand{\bJ} {{\bf J}}
\newcommand{\bL} {{\bf L}}
\newcommand{\be} {{\bf e}}
\newcommand{\betat}{\tilde{\beta}}

\newcommand{\dss}{D^{\ast \ast}}

\newcommand{\at}{a_{2}(1320)}
\newcommand{\ao}{a_{1}(1260)}

\newcommand{\bo}{b_{1}(1235)}
\newcommand{\ft}{f_{2}(1270)}
\newcommand{\fo}{f_{1}(1285)}
\newcommand{\fz}{f_{0}(1300)}
\newcommand{\ho}{h_{1}(1170)}
\newcommand{\kt}{K_{2}^{\ast}(1430)}
\newcommand{\kl}{K_{1}(1270)}
\newcommand{\kz}{K_{0}^{\ast}(1430)}
\newcommand{\kh}{K_{1}(1400)}

\newcommand{\eqntimes}{\mbox{} \times}
\newcommand{\eqnhspace}{\hspace{3em}}
\newcommand{\intereqnvspace}{\vspace{-1ex}}
\newcommand{\eqnneghspace}{\hspace{-6ex}}
\newcommand{\beqn}{\begin{equation}}
\newcommand{\eeqn}{\end{equation}}
\newcommand{\barr}[1]{\begin{array}{#1}}
\newcommand{\earr}{\end{array}}
\newcommand{\beqna}{\begin{eqnarray}}
\newcommand{\eeqna}{\end{eqnarray}}
\newcommand{\btablec}{\begin{table} \begin{center}}
\newcommand{\etablec}{\end{center} \end{table}}

\begin{document}
\title{\small \rm \begin{flushright} \small{hep-ph/9411301}\\
\small{RAL-94-116}\\
\small{OUTP-94-29P} \end{flushright} \vspace{2cm}
\LARGE \bf The production and decay of hybrid mesons
 by flux-tube breaking \vspace{0.8cm} }
\author{Frank E. Close\thanks{E-mail : fec@v2.rl.ac.uk} \\
{\small \em Particle Theory, Rutherford-Appleton Laboratory, Chilton,
Didcot OX11 0QX, UK} \\ \\
Philip R. Page\thanks{E-mail : p.page@physics.oxford.ac.uk} \\
{\small \em Theoretical Physics, University of Oxford, 1 Keble Road, Oxford OX1
3NP, UK}  \\  \\}
\date{November 1994 \vspace{1.5cm}}

\begin{center}
\maketitle

\begin{abstract}

An analytic calculation of the breaking of excited chromoelectric
flux-tubes is performed in an harmonic oscillator approximation and
applied to predict the dynamics of all  $J^{PC}$ low-lying
gluonic excitations of mesons (hybrids).
Widths, branching ratios and production dynamics of some
recently discovered $J^{PC} = 1^{-+}, 0^{-+}$ and $1^{--}$ mesons are found
to be in remarkable agreement with these results. We introduce the
selection rules that can be used to understand the systematics of
numerical decay calculations and we find possible significant breaking
of these rules for specific channels that may enable enhanced
production and detection of hybrids.

\end{abstract}
\end{center}

\newpage

\section{Introduction}

Alongside justifiable pride in establishing and applying the standard
model, we should also recognise that there remains an area of substantial
fundamental ignorance: while the gluon degrees of freedom expressed
in $L_{QCD}$ have been established beyond doubt in high momentum data,
their dynamics in the strongly interacting limit epitomised by hadron
spectroscopy are quite obscure.
It is possible that this is about to change as candidates for
gluonic hadrons (glueballs and hybrids) are now emerging \cite{amsler94}.

 For the first time there is a  candidate scalar glueball \cite{kirk,anisov}
 whose mass
$1.5 \sim 1.6$ GeV is consistent with the prediction of $1550 \pm 50$ MeV from
lattice studies of QCD \cite{lattice}. Simulations of the
lattice dynamics, where gluonic fields are modelled as flux-tubes,
reproduce these numbers for glueballs \cite{paton85} and predict that hybrid
mesons will be manifested in the $1.5 - 2 $ GeV mass
range \cite{paton85,swanson94}. This is where candidates are now emerging,
in particular the predicted \cite{barnes83,sharpe83,paton85}
family of lowest lying multiplets of
hybrid hadrons including
$J^{PC}={(0,1,2)}^{-+},1^{--}$. Of these states,
potentially the most clear cut
as a hybrid would be the $J^{PC} = 1^{-+}$
exotic  in data from
the AGS at Brookhaven with mass of about 2 GeV
\cite{lee94}. Less unique signatures but nonetheless potential
$0^{-+}$ and $2^{-+}$ partners
are seen in this same mass region in diffractive production
by the VES Collaboration at Serpukhov \cite{ves,close94}.
Extensive and thorough analyses of the $1^{--}$ system, which is especially
well probed experimentally due to its isolation in $e^+e^-$ annihilation
and photoproduction, show that in the 1.4 to 2 GeV range of interest ``mixing
with non-$q\bar{q}$ states must occur" \cite{don2}.
 If lattice simulations \cite{perantonis90} and the modelling of hybrid
decays are
reliable, the production and decay of
charmonium hybrids at the Tevatron may be responsible, in part, for
the anomalous production of $\psi$ and $\psi'$ observed at CDF \cite{closecdf}.

Predictions for masses and/or the $J^{PC}$ patterns of multiplets
are a guide to identifying potential candidates for hybrids and
glueballs but alone will not establish the spectroscopy and dynamics.
Characteristic production and decay signatures proved seminal in
establishing the light quark $\bar{Q}Q$ nonets and will be no less important
in exciting and recognising gluonic hadrons.
While the masses of glueballs and hybrids are computable
in lattice QCD, at least in the quenched approximation, the decay
dynamics are at present beyond its reach. However,
intuition gained from the strong coupling expansion of lattice QCD has
inspired the development of flux-tube models of mesons, which are
probably the nearest we have to a realistic simulation of strong gluon
dynamics, whereby the
decay amplitudes may be computed \cite{kokoski87}. In this picture
mesons consist of $Q\bar{Q}$ connected by
a cylindrical bag of coloured fields: the ``flux-tube". When the
flux tube is in its ground state, the excitation of the $Q\bar{Q}$
degree of freedom yields the conventional meson spectrum; excited modes of
the flux-tube are also natural in strong QCD and this leads to a set
of states that have yet to be confirmed by experiment. It is the existence
of these ``hybrid" states (where the flux-tube is excited or ``plucked"
in the presence of the $Q\bar{Q}$ coloured sources) that remains an
open question within QCD dynamics and, as such, a missing part of the
standard model.

 Isgur,  Paton and collaborators have developed and applied this model with
some success to the decays of conventional
mesons \cite{kokoski87} and have also given
some limited predictions for the decays of a few hybrid states
\cite{kokoski85},
specifically those with light flavours and exotic $J^{PC}$ quantum numbers.
It is the latter that have in part motivated the search strategy
for hybrids \cite{lee94,cernproposal}. This was adopted by the BNL experiment
\cite{lee94}
who have studied the production of $\pi f_1(1285)$ and find that
 over 40\% of the signal is in the exotic $1^{-+}$ partial wave;
however the production mechanism appears to involve $\rho$ exchange and is
in sharp contrast to the expectations.

It is therefore timely to examine in more detail the implications
of the model and the experimental signatures.

A powerful and empirically successful approach \cite{kokoski87,isgur89},
has been to use S.H.O. basis wavefunctions, thereby enabling analytic studies
that reveal the relationships among amplitudes. This has been
employed
for the flux-tube model in the case of the decays of conventional mesons
\cite{kokoski87} and the point of departure for the present paper will
be to make an analogous application to the dynamics of hybrid mesons. We
find that this analytic approximation

\begin{enumerate}

\item reproduces the numerical results of Isgur, Kokoski and Paton
\cite{kokoski85} with rather good accuracy,

\item reveals for the first time the relationships that exist among
amplitudes and which underpin their relative
magnitudes, thereby highlighting signals
that are potentially significant
on  general grounds rather than due to specific choices
of parameters

\item exhibits the explicit dependence of amplitudes and widths on masses and
other parameters thereby enabling application to possible
candidates as they emerge, in particular the ${(0,1,2)}^{-+}$ states
recently reported from VES and BNL \cite{ves,lee94} and
is  immediately  extendable to heavy quark hybrids

\item provides the first detailed analysis of hybrid decays for both light
and heavy flavours, for arbitrary relevant quantum numbers.

\end{enumerate}

We find that the possibility of the
VES $0^{-+}$ being a hybrid \cite{close94} is strongly supported by our
analysis
whereas the VES $2^{-+}$ state is less clear. The $1^{--}$ partial wave
shows clear signals consistent with hybrid excitation in the branching
ratios of $\rho(1460)$ \cite{don2}. Furthermore
we find  that the selection rule that hybrids do not
decay into ground state mesons may in some cases
(notably $\pi \rho$ decays) be significantly
broken. This may explain the production mechanism of the $1^{-+}$ candidate
 at BNL \cite{lee94} and
it also suggests new ways of accessing the hybrid sector.

                Having obtained analytic closed forms for the decay amplitudes
it is possible to apply these results to the dynamics of heavy hybrids, in
particular hybrid charmonium.
Heavy hybrids are specifically interesting, because heavy
quarkonium is well understood,
and the excitations of gluonic degrees of freedom for charmonium
are predicted to be manifested
in the vicinity of $DD \sim DD^{**}$
charm thresholds \cite{swanson94} and so should be clean
experimentally. We shall report on this elsewhere \cite{page3}
as the numerical results depend in part on parameters that need first
to be confronted with the dynamics of the conventional charmonia
for which there is no analogous discussion in the literature known to
us \cite{page2}. In the present paper we shall restrict our attention
primarily to the light flavoured hybrids, exposing the analytic and parameter
dependence of the decay amplitudes, and both updating and extending
the  numerical
studies of ref. \cite{kokoski85} in the light of modern data.
As already mentioned, this will expose the dynamical origin of the
familiar selection rules for hybrid decays and, more important, reveal
significant violations of them for certain decays or production
channels.

The structure of the paper is as follows.

In section 2 we present the general formalism describing mesons in
flux-tube models, for both conventional and hybrid configurations.
This leads to the master equation (eqn. \ref{flux28943}) controlling the
structure of decay amplitudes. Section 3 shows how this leads to
selection rules, in particular that the dominant two-body decays
of the lowest hybrids are into $L=0$ and $L=1$ (``S + P") $q\bar{q}$
meson pairs. These dominant channels are studied in some detail
in section 4 where we extract the analytic structures, reproduce
existing computations for exotic $J^{PC}$ hybrids and extend them to
all low-lying
$J^{PC} (J \leq 2)$ combinations for both light and heavy flavours.
In section 5 we investigate significant violations of the selection
rules, in particular noting that the channels $0^{-+},1^{++},1^{-+},
1^{--} \rightarrow \pi + \rho, \; \pi + \gamma,
\; KK^*, \; \pi\eta$ may be non-negligible.
In the conclusions, section 6, we confront these results with emerging
candidates, highlighting the positive and negative features of the
VES and BNL ${(0,1,2)}^{-+}$ states, and consider the implications for
diffractive photoproduction of vector hybrids and the future
extension to hybrid charmonium.


\section{General  formalism for mesons and their decays \label{section2}}

The flux-tube model was motivated by the strong coupling expansion of lattice
QCD and to some extent by early descriptions of flux-tubes as cylindrical
bags of coloured fields \cite{paton85,horn,hasenfratz}.
 As application of plaquette operators
in lattice QCD extends a line of flux only in the transverse direction, then
in the flux-tube model by analogy one allows only locally transverse
fluctuations of the tube.
There are correspondingly {\it two} degrees of freedom for excitation of
mesons: the relative coordinate  $\br_{A} \equiv \bar{\br}-\br$ of the
$\bar{Q}-Q$ and the transverse coordinate of the flux-tube
 $\by_{\perp} \equiv -(\by \times \hat{\br}_{A})
\times \hat{\br}_{A}$ (see fig. 4a of ref. \cite{kokoski87}).

The flux-tube which connects the quark and antiquark may be
 represented by a system of $N$
beads a distance $a$ apart in modes $\{ n_{m+},n_{m-} \}$
\cite{paton85} ,
vibrating w.r.t.\ the $Q\bar{Q}$- axis $\br_{A}$ as equilibrium position.
The beads are connected to each other and to the quarks at the ends via
a non-relativistic string potential with string tension $b$.
The flux-tube has in general $\Lambda$
units of angular momentum around the Q\={Q}-axis.

Decay occurs when the flux-tube breaks
at any point along its length, producing in the process
a $\bar{q}q$ pair in a relative $J^{PC}=0^{++}$ state. This is
similar in spirit to the old Quark Pair Creation or $^3P_0$ model \cite{pene}
 but with an essential difference.
In the $^3P_0$ model the $\bar{q}q$ have equal chance of being created
at any distance from the initial $\bar{Q}Q$ axis (the ``tube" is infinitely
thick) whereas in the flux-tube model
 the  distribution of the $^3P_0$ pair {\it transverse} to the $\bar{Q}Q$
axis is
controlled by the transverse ($\by_{\perp}$) distribution of the flux-tube.
For conventional mesons, where the flux-tube is in
its ground state, this distribution is parametrised as a Gaussian
(eqn. \ref{mesondec} below) such that the tube has a finite width; for hybrid
states where the string is excited, the distribution is more
structured, in particular containing a node along the initial
$\bar{Q}Q$ axis (eqn. \ref{hybriddec} below).
This gives characteristic constraints
on hybrid decay amplitudes, in particular leading to a selection
rule against certain hybrid decay modes.

Specifically, the pair creation amplitudes are formulated as follows.

The pair creation position $\by$ \cite{kokoski87} is measured relative
to the origin
(the CM of the initial quarks) and
 $\by_{\perp} \equiv -(\by \times \hat{\br}_{A})
\times \hat{\br}_{A}$.
The Q\={q} - axes of the final states B and C
are $\br_{B} = \br_{A}/2 + \by$ and $\br_{C} = \br_{A}/2 - \by$ respectively.
The flux-tube overlap $\gamma (\br_{A},\by_{\perp})$ (defined below)
 is assumed to be independent  of
$\by_{\parallel} \equiv (\by \cdot \hat{\br}_{A}) \hat{\br}_{A}$
 \cite{kokoski87} .

When the flux-tube is in its ground state ({\it conventional mesons}) the
pair creation amplitude is

\beqn
\label{mesondec}
\gamma (\br_{A},\by_{\perp}) = A^{0}_{00} \sqrt{\frac{fb}{\pi}}
 \exp (-\frac{fb}{2} \by^{2}_{\perp})
\eeqn
The thickness of the flux tube is related inversely to $f$
(where the infinitely thick flux tube with $f=0$ corresponds to the
$^{3}P_{0}$ -
model). A detailed discussion of these quantities and the structure of
eqn. \ref{mesondec} may be found in ref. \cite[eqn. (A21)]{kokoski87}
and ref. \cite{perantonis87}.

In this work we consider the decay of the
energetically lowest lying {\em hybrid meson} A
with
$\{n_{m+},n_{m-}\} =
 (n_{1 +}=1 \, or \, n_{1 -} = 1 \, , \, n_{m \pm}=0 \; \forall \, m \neq 1)$
and $\Lambda_{A} = \pm 1$ \cite{paton85}.
The flux-tube overlap \cite{perantonis87,kokoski85} prohibits pair
creation on the hybrid Q\={Q}-axis :

\beqn
\label{hybriddec}
\gamma (\br_{A},\by_{\perp}) = \kappa \sqrt{2b}
A^{0}_{00} \sqrt{\frac{fb}{\pi}}
\be_{\Lambda}(\hat{\br}_{A}) \cdot \by_{\perp}
\exp (-\frac{fb}{2} \by^{2}_{\perp})
\eeqn

Here $\kappa$ is approximately
constant \cite{perantonis87}
and $\be_{\Lambda}(\hat{\br}_{A})$ refers to body-fixed coordinate system.

The entire analysis follows once the full  amplitude for the process
$A \rightarrow BC$  has been
formulated. The master equation is eqn. \ref{flux28943} below.
 Its qualitative structure
involves the wavefunctions $\psi(\vec{r})$ for the $\bar{Q}Q$ degrees of
freedom of the mesons \{$A;BC$\}, the $\gamma(\br_{A},\by_{\perp})$ being the
flux-tube breaking amplitude and the $\bsigma$ and $\bnabla$ factors
reflecting the $^3P_0$ quantum numbers of the created pair.

To specify the decay amplitude,
consider a quark-antiquark bound system A with quark at position $\br$
and antiquark
at $\bar{\br}$. The system has a
momentum $\bp_{A}$ and wavefunction
 $\psi_{A}^{LM_{L} \Lambda} (\bar{\br}-\br)$, with
relative coordinate $\br_{A} \equiv \bar{\br}-\br$,
angular momentum $L_{A}$ and projection $M_{L}^{A}$.

We shall focus on two initial
quarks of mass $M$, with pair creation of
quarks of mass $m$. The decay amplitude $M(A \rightarrow B C)
\equiv \langle A \! \mid \hat{\cal C}
\mid \! B C \rangle$ as obtained from eqns. \ref{flux83624} and
\ref{flux83625} (Appendix A) can be shown, in
the rest frame of A,
 to be given by

\beqna
\label{flux28943}
\lefteqn{\eqnneghspace
M^{M_{L}^{A}M_{L}^{B}M_{L}^{C}}_{M_{S}^{A}M_{S}^{B}M_{S}^{C}}(A
\rightarrow B C) =
- \frac{a \tilde{c}}{9 \sqrt{3}} (2 \pi)^{3}
\delta^{3} (\bp_{B} + \bp_{C}) \frac{i}{2} Tr(A^{T}BC)^{F}
Tr(A^{T}B \bsigma^{T} C)^{SM_{S}} \cdot \nonumber } \\
& &\eqnneghspace \eqntimes \int d^{3} \br_{A} \, d^{3} \by \,
\psi_{A}^{LM_{L} \Lambda}(\br_{A}) \exp (i \frac{M}{m+M} \bp_{B}
\cdot \br_{A}) \gamma (\br_{A},\by_{\perp}) \nonumber \\
& & \eqnneghspace \eqntimes (i \bnabla_{\br_{B}} + i \bnabla_{\br_{C}} +
\frac{2 m}{m+M} \bp_{B})
{\psi_{B}^{LM_{L} \Lambda}}^{\ast}(\br_{B})
{\psi_{C}^{LM_{L} \Lambda}}^{\ast}(\br_{C}) \; + \; (B \leftrightarrow C)
\eeqna
where $a\tilde{c}$ is an unknown constant, subsumed into the quantity
$\frac{a \tilde{c}}{9 \sqrt{3}} \frac{1}{2} A^{0}_{00} \sqrt{\frac{fb}{\pi}}$
(see Appendix A and also eqn. \ref{flux19731} later) that has been fitted
\cite{kokoski87}
to the known decays of ordinary mesons. Since the lattice spacing $a$
sets the scale in the flux-tube model, it is introduced to
obtain the correct dimension for $M(A \rightarrow B C)$.

Also $(B \leftrightarrow C)$ indicates a term obtained by interchanging flavour
$B^{F} \leftrightarrow C^{F}$, spin $B^{SM_{S}} \leftrightarrow C^{SM_{S}}$
and momenta $\bp_{B} \leftrightarrow \bp_{C}$ in the first term in
eqn. \ref{flux28943}. We shall refer to the last two lines in eqn.
\ref{flux28943} as the {\em space part} of $M(A \rightarrow B C)$. This
is the same for both terms up to an overall sign and so
from now on it is sufficient just to consider the exhibited term.

The helicity amplitude can now be constructed from the L-S basis amplitude

\beqna
\label{flux93518}
\lefteqn{M_{M_{J}^{A} M_{J}^{B} M_{J}^{C}}(A \rightarrow B C) =
\sum_{ { \{ M_{L} , M_{S} \} }^{A,B,C} }
M^{M_{L}^{A} M_{L}^{B} M_{L}^{C}}_{M_{S}^{A} M_{S}^{B }M_{S}^{C}}(A
\rightarrow B C)} \nonumber \\
& & \eqntimes \langle L_{A}M_{L}^{A} \, S_{A}M_{S}^{A} \! \mid \!
J_{A}M_{J}^{A} \rangle \:
\langle L_{B}M_{L}^{B} \, S_{B}M_{S}^{B} \! \mid \! J_{B}M_{J}^{B} \rangle \:
\langle L_{C}M_{L}^{C} \, S_{C}M_{S}^{C} \! \mid \! J_{C}M_{J}^{C} \rangle \:
\eeqna

To convert to partial wave amplitudes we perform
 the vector sum of the three total angular
momenta $\bJ_{A}$ , $\bJ_{B}$ and $\bJ_{C}$ in the order
$\bJ = \bJ_{B} + \bJ_{C}$ and $\bJ_{A} = \bJ + \bL$ , and obtain the
Jacob-Wick formula \cite{jacob59}

\beqna
\label{flux98321}
M_{LJ}(A \rightarrow B C) & = & \sqrt{\frac{2L+1}{2J_{A}+1}}
\sum_{{ \{ M_{J} \} }^{A,B,C}}
M_{M_{J}^{A} M_{J}^{B} M_{J}^{C}}(A \rightarrow B C) \nonumber \\ & &
\times \: \langle L0 \, JM_{J}^{A} \mid J_{A}M_{J}^{A} \rangle \:
\langle J_{B}M_{J}^{B} \, J_{C}M_{J}^{C} \mid JM_{J}^{A} \rangle
\eeqna

The decay rate $\Gamma_{LJ}(A \rightarrow B C)$ is calculated according
to the prescription of Isgur and Kokoski \cite{kokoski87}. Here A, B and C
are assumed to be narrow resonances obeying non-relativistic kinematics,
while some relativistic effects are taken into account by using
masses $\tilde{M}$ as defined in ref. \cite{kokoski87} :

\beqn
\Gamma_{LJ}(A \rightarrow B C) = \frac{p_{B}}{(2 J_{A} + 1) \pi}
\frac{\tilde{M}_{B} \tilde{M}_{C}}{\tilde{M}_{A}} \;
{\mid M_{LJ}(A \rightarrow B C) \mid}^{2}
\eeqn

Throughout this work, all resonances are assumed to be narrow, and threshold
effects are not taken into account.

The decay amplitudes and widths may now be calculated once the wavefunctions
$\psi(\vec{r})$ and values of parameters are specified. We shall apply these
calculations to a broad class of hybrid decays: the ``allowed" couplings
to two-body final states consisting of $L=0$ and $L=1$ $Q\bar{q}$
states and the ``forbidden" transitions where both of the produced
mesons
are $L=0$ $Q\bar{q}$-states. In the former class
we shall recover the numerical results of Isgur et al. \cite{kokoski87}
as a particular
case. Our analytical results enable extension beyond previous works, in
particular to the class of forbidden decays some of which, as we shall
see, may be significant and hence offer the prospect of enhanced production.


\section{Selection Rules for Hybrid Mesons \label{section3} }

The literature contains detailed studies of the decays of {\it ordinary mesons}
both numerically with exact wavefunctions appropriate to a QCD-improved quark
model \cite{isgur85} and analytically in an harmonic oscillator
approximation \cite{kokoski87}. For the case of {\it hybrid} mesons the
rather limited literature reports only numerical results and
for a restricted class of ``exotic" $J^{PC}$ only \cite{kokoski85}.
As noted in ref. \cite{kokoski87} the analytic forms reveal the
relationships that exist among amplitudes as well as establishing the
sensitivity of results to parameters. The consistency of the results
in the two approaches for the ordinary mesons encourages us to
 perform an analogous {\it analytical} calculation of the decay amplitude
of an initial energetically lowest lying {\it hybrid meson}
 A having $L_{A}=1$, with outgoing ordinary mesons B and C
($\Lambda_{B} = \Lambda_{C} = 0$) having $L_{B}=$ 0 or 1 and $L_{C}=0$.
We employ S.H.O.\ wave functions with inverse radii $\beta_{B}$ and
$\beta_{C}$ for the corresponding states

\[
\psi_{C}^{L=0}(\br) = {\cal N}_{C} \exp (- {\beta_{C}^{2}r^{2}}/2)
\eqnhspace\psi_{B}^{L=1}(\br) = {\cal N}_{B} r Y_{1 M_{L}^{B}} (\hat{\br})
\exp (- {\beta_{B}^{2}r^{2}}/2)
\]

\intereqnvspace
\beqn
\label{flux94294}
{\cal N}_{C} = \frac{\beta_{C}^{3/2}}{\pi^{3/4}}
\eqnhspace{\cal N}_{B} = 2 \sqrt{\frac{2}{3}} \frac{\beta_{B}^{5/2}}{\pi^{1/4}}
\eeqn

For the case of $B,C$ both $L=0$, the $\psi_C(\br)$ is to be used with
$\beta_C \rightarrow \beta_B$ and ${\cal N}_{C} \rightarrow {\cal
N}_{B}$ as necessary.

For hybrids there is a centrifugal barrier for the $Q\bar{Q}$
pair that arises from the matrix element of $\vec{L}_Q^2$ in the
full quark-and-flux-tube angular momentum eigenstate. The angular wavefunction
of the combined gluon or flux-tube and quark system was discussed by Horn and
Mandula \cite{horn} and subsequently by Hasenfratz
{\it et.al.} \cite{hasenfratz} and by Isgur and Paton \cite{isgur85}. The
latter references give essentially the same rigid body angular wavefunction
for the full system, which is in the {\it body-fixed} coordinate system

\beqn
 \psi_{A}^{LM_{L}^{A} \Lambda}(\br_{A}) \sim {\cal D}^{1}_{M_{L}^{A}
\Lambda}(\phi,\theta,-\phi)
\eeqn

This is the amplitude given by Isgur and Paton \cite{paton85}
to find the $Q\bar{Q}$-axis pointing along
$(\theta,\phi)$, in a hybrid state with total orbital angular momentum
$L_{A}$ and $z$-projection $M_{L}^{A}$
for a flux-tube with $\Lambda$ units of angular momentum around the
$Q\bar{Q}$-axis. We shall restrict attention to the
lowest lying state where $L_{A} = 1, \Lambda = \pm 1$.

We allow for a general radial dependence of the hybrid wave function
parameterized by $\delta$, with $0<\delta \leq 1$,

\beqn
\label{flux82045}
\psi_{A}(\br) = {\cal N}_{A} r^{\delta} {\cal D}^{1}_{M_{L}^{A}
 \Lambda}(\phi,\theta,-\phi) \exp (- {\beta_{A}^{2}r^{2}}/2)
\eqnhspace {\cal N}_{A} = \sqrt{\frac{3 \beta_{A}^{3 +
 2 \delta}}{2 \pi \Gamma(3/2+\delta)}}
\eeqn

Isgur and Paton \cite[eqn. 28]{paton85} introduced a simple
approximation for the matrix element of $\vec{L}_Q^2$ in this state,
which neglects a mixing term that raises and lowers $\Lambda$. This
approximation gives $ \vec{L}_Q^2 = L_{A} (L_{A} + 1) - \Lambda^{2}$
which transforms the Schrodinger equation into an ordinary differential
equation
for the (adiabatic) radial wavefunction. Thus, for our case where
${L}_A = 1, \Lambda = \pm 1$ we can reproduce the small r behaviour
of the Schr\"{o}dinger equation for the hybrid meson
 by choosing $\delta = 0.62$, satisfying $\delta (\delta + 1) =
L_{A} (L_{A} + 1) - \Lambda^{2} = 1$. (In practice
the values $\delta = 0.62$ or 1 give similar numerical results).

The essential origin of the much advertised selection rules and
their violation is driven by the third line of the master equation
\ref{flux28943} as we shall now see.

 For the case of B and C  being $L=0$ $q\bar{q}$ mesons
where both have the wave function $\psi_{C}(\br)$ in eqn. \ref{flux94294},
but with
$\beta_{B} \neq \beta_{C}$ in general, one has

\beqna
\label{flux81639}
\lefteqn{ \be^{\ast}_{\sigma} \cdot (i \bnabla_{\br_{B}}
+ i \bnabla_{\br_{C}} + \frac{2 m}{m+M}
\bp_{B}) \: \psi_{B}^{\ast}(\br_{B}) \psi_{C}^{\ast}(\br_{C})
= } \nonumber \\  & & {\cal N}_{B} {\cal N}_{C} \exp (-\tilde{\beta}^{2}
(\frac{\br_{A}^{2}}{4}+\by^{2}) - \frac{\Delta}{2} \br_{A} \cdot \by) \;
\be^{\ast}_{\sigma} \cdot (- i \tilde{\beta}^{2} \br_{A} - i \Delta \by +
\frac{2 m}{m+M} \bp_{B} )
\eeqna
with the average $\tilde{\beta}^{2} \equiv (\beta_{B}^{2}+\beta_{C}^{2})/2$
and difference $\Delta \equiv \beta_{B}^{2}-\beta_{C}^{2}$.
 The nature of the selection rule suppressing the transition to $L=0$ states
 arises when we perform the y-integration for which only terms linear in
{\bf y} contribute (essentially due to the $\by_{\perp}$
 factor in the pair creation amplitude
 $ \gamma (\br_{A},\by_{\perp})$
eqn. \ref{hybriddec}), and hence the result is linearly proportional
to $\Delta$, the multiplier of of these terms. To the extent that hadrons have
the same size, such that $\beta_{B} = \beta_{C}$, the integral vanishes and
hence the selection rule is immediate
(we shall consider the corrections due to
$\beta_{B} \neq \beta_{C}$ in \S \ref{section5}).

By contrast, for the case where  the B+C system consists of
a $L=0$ and $L=1 \; Q\bar{q}$ meson pair
 the corresponding expression becomes

\beqna
\label{flux76283}
\lefteqn{(i \bnabla_{\br_{B}} + i \bnabla_{\br_{C}} + \frac{2 m}{m+M}
\bp_{B}) \:
\psi_{B}^{\ast}(\br_{B}) \psi_{C}^{\ast}(\br_{C})
= {\cal N}_{C} \exp (-\tilde{\beta}^{2}
(\frac{\br_{A}^{2}}{4}+\by^{2}) -
\frac{\Delta}{2} \br_{A} \cdot \by) } \nonumber \\
& & \eqntimes {\cal N}_{B} [i \sqrt{\frac{3}{4 \pi}}
\be^{\ast}_{M_{L}^{B}} +
(\frac{r_{A}}{2} Y_{1 M_{L}^{B}}^{\ast} (\hat{\br}_{A})
+ y Y_{1 M_{L}^{B}}^{\ast} (\hat{\by}) ) \:
(- i \tilde{\beta}^{2} \br_{A} - i \Delta \by + \frac{2 m}{m+M} \bp_{B} )]
\eeqna

The approximation
of equal size gives the leading non-vanishing amplitude in general
and allowing   $\beta_{B} \neq \beta_{C}$ gives corrections. In this
first orientation we shall simplify to the approximation
 $\beta_{B} = \beta_{C}$ whence

\beqna
\label{flux20732}
\lefteqn{ \eqnneghspace \eqnneghspace  \int d^{3} \by \gamma
(\br_{A},\by_{\perp}) \times (eqn. \,
\ref{flux76283}) =
\sqrt{\frac{3}{8 \pi}} {\cal N}_{S}
{\cal N}_{B} {\cal N}_{C}(-i \tilde{\beta}^{2} \br_{A}
+ \frac{2 m}{m+M} \bp_{B})
} \nonumber \\ & & \eqnneghspace
\eqntimes \exp (- \frac{\tilde{\beta}^{2} \br_{A}^{2}}{4})
\sum_{M_{L}^{'B}} {\cal D}^{1 \: \ast}_{M_{L}^{B} M_{L}^{'B}}
(\phi,\theta,-\phi) \bar{\gamma}_{M_{L}^{'B} \Lambda}^{1} \nonumber
\eeqna
\intereqnvspace
\beqn
\bar{\gamma}_{M_{L}^{'B} \Lambda}^{1}  \equiv
2 \int d^{3} \by \; \be_{\Lambda}(\hat{\br}_{A}) \cdot \by_{\perp} \;
\be_{M_{L}^{'B}}^{\ast} \cdot \by \;
\exp (-\tilde{\beta}^{2} \by^{2} - \frac{fb}{2} \by^{2}_{\perp})
\eeqn
where we used $y Y_{1 M_{L}^{'B}}^{\ast} (\hat{\by}) =
\sqrt{\frac{3}{4 \pi}} \be_{M_{L}^{'B}}^{\ast} \cdot \by$, and defined
${\cal N}_{S} \equiv \kappa \sqrt{b} A^{0}_{00} \sqrt{\frac{fb}{\pi} }$.

The angular momentum projection $M_{L}^{B}$ is defined relative to
the {\it space-fixed axes} (with $\bp_{B}$ defining the $\hat{\bz}$-axis),
as usual. The y-integration is done in the system of
{\it body-fixed axes}  (with the q\={q}-axis defining the $\hat{\bz}$-axis)
and so we must convert to angular momentum projection $M_{L}^{'B}$
relative to the body-fixed system.
The body is moving with
its $\hat{\bz}$-axis rotated by rotation matrix
$\cal R$ relative to the space-fixed
coordinate system, i.e.\ $\mid \! \psi (\hat{\by}_{\br}) \rangle
= {\cal R} \mid \! \psi (\hat{\by}_{r \hat{\bz}}) \rangle$.
The spherical harmonics transform as

\[
Y_{L M} (\hat{\by}_{\br}) =
\langle \psi (\hat{\by}_{\br}) \! \mid \! \psi (r L M) \rangle  =
\sum_{M^{'}} \langle \psi (\hat{\by}_{r \hat{\bz}}) \! \mid
\! \psi (r L M^{'}) \rangle
\langle \psi (r L M^{'}) \! \mid {\cal R}^{+} \mid \! \psi (r L M)
\rangle
\]
\intereqnvspace
\beqn
= \sum_{M^{'}} Y_{L M^{'}} (\hat{\by}_{r \hat{\bz}})
{\cal D}^{L}_{M M^{'}} (\phi,\theta,-\phi)
\eeqn
where we used ${\cal D}^{L}_{M M^{'}} (\phi,\theta,-\phi) \equiv
\langle \psi (r L M^{'}) \! \mid {\cal R}^{+} \mid \! \psi (r L M) \rangle$
\cite[Appendix A]{paton85}.
Performing the $\by$ integration in eqn. \ref{flux20732}

\beqn
\label{flux20416}
\bar{\gamma}_{M_{L}^{'B} \Lambda}^{1}   = \frac{\pi^{3/2}}{\betat
{(\betat^{2} + f b / 2)}^{2}} \delta_{M_{L}^{'B} \Lambda} \equiv
\bar{\gamma}^{1} \: \delta_{M_{L}^{'B} \Lambda}
\eeqn
and hence there is   an important selection
rule operating in the moving frame of the initial $Q\bar{Q}$-pair : The
one unit of angular momentum of the incoming hybrid
around its $Q\bar{Q}$-axis is exactly absorbed by the component of the
angular momentum of the outgoing meson B along this axis.
This helps to generate relationships among amplitudes
for decays into $L=0$ and $L=1 \; Q\bar{q}$ states.


\section{Hybrid meson decay into $L=1$ and pseudoscalar
mesons \label{flux92712} }

The overall strengths follow once integration over $\br_{A}$ is performed.
In the harmonic oscillator basis the amplitudes can be calculated
analytically or at least reduced to  tractable forms
that expose their detailed structure and parametric dependences.

Expanding $\exp (i \frac{M}{m+M} \bp_{B} \cdot \br_{A})$ in
spherical Bessel functions and Legendre polynomials
we obtain for the space part of the decay
amplitude in eqn. \ref{flux28943} (using eqns. \ref{flux20732} and
\ref{flux20416})

\beqna
\lefteqn{\eqnneghspace M(A \rightarrow B C) =
\sqrt{\frac {3}{8 \pi}}
{\cal N}_{A} {\cal N}_{B} {\cal N}_{C} {\cal N}_{S} \bar{\gamma}^{1}
\sum_{n=0}^{\infty} (2 n + 1) i^{n}} \nonumber \\ & & \eqnneghspace \eqntimes
\int_{0}^{\infty} dr_{A} r_{A}^{2+\delta} j_{n}(\frac{M}{m+M} p_{B}r_{A})
\exp (- (2 \beta_{A}^{2} + \betat^{2}) \frac{r_{A}^{2}}{4}) \nonumber \\
& & \eqnneghspace \eqntimes \int d \Omega_{\hat{\br}_{A}} (- i \betat^{2}
\br_{A} +
\frac{2 m}{m+M} \bp_{B})
{\cal D}^{1}_{M_{L}^{A} \Lambda} (\phi,\theta,-\phi)
{\cal D}^{1 \: \ast}_{M_{L}^{B} \Lambda} (\phi,\theta,-\phi)
P_{n}(cos \theta)
\eeqna

The partial wave amplitudes $M_{L} (A \rightarrow BC)$
(with $J = J_{B}$) of eqn. \ref{flux98321}
can now be evaluated. We obtain (modulo the factor
${(2 \pi)}^{3} \delta^{3} (\bp_{B} + \bp_{C})$)
\beqn
\label{flux19731}
M_{L} (A \rightarrow B C) = (\frac{a \tilde{c}}{9 \sqrt{3}}
\frac{1}{2} A^{0}_{00} \sqrt{\frac{fb}{\pi}})
\frac{\kappa \sqrt{b}}{{(1 + f b / (2 \betat^{2}) \, )}^{2}}
\sqrt{\frac{2 \pi}{3 \Gamma(3/2+\delta)}}
\frac{\beta_{A}^{3/2 + \delta}}{\betat} \tilde{M}_{L} (A \rightarrow B
C)
\eeqn

In Table \ref{flux75423} we display the reduced partial wave amplitudes
$\tilde{M}_{L} (A \rightarrow BC)$
in a compact form by defining

\beqn \begin{array}{lll} \label{flux84610}
S = -(3 \tilde{h}_{0} - \tilde{g}_{1} + 4 \tilde{h}_{2}) &
P_{1} = -i (2 \tilde{g}_{0} + 3 \tilde{h}_{1} - \tilde{g}_{2}) &
P_{2} = -i (\tilde{g}_{0} + \tilde{g}_{2})                      \\
P_{3} = -i (5 \tilde{g}_{0} + 3 \tilde{h}_{1} + 2 \tilde{g}_{2})  &
P_{4} = -i (10 \tilde{g}_{0} + 9 \tilde{h}_{1} + \tilde{g}_{2})   &
P_{5} = -i (5 \tilde{g}_{0} + 6 \tilde{h}_{1} - \tilde{g}_{2})   \\
D = (\tilde{g}_{1} + 5 \tilde{h}_{2})                             &
F = -3 i (\tilde{g}_{2} + \tilde{h}_{3})                          &
G = 0
\end{array} \eeqn

\beqn
\label{flux75422}
 \left( \barr{c} \tilde{g}_{n} \\ \tilde{h}_{n}  \earr \right)
= \left( \barr{c}  \frac{2 m}{m+M} \bp_{B} \\ \betat^{2}  \earr \right)
\int_{0}^{\infty} dr
\left( \barr{c}  1 \\ r  \earr \right)
r^{2+\delta} j_{n}(\frac{M}{m+M} p_{B}r)
\exp (- (2 \beta_{A}^{2} + \betat^{2}) \frac{r^{2}}{4})
\eeqn

For $\delta = 1$ explicit evaluation of $\tilde{g}_{1}$, $\tilde{h}_{0}$
and $\tilde{h}_{2}$ can be made using

\beqn
\label{flux82643}
\int_{0}^{\infty} dr
\left( \barr{c}  j_{1} (u r) \\ r j_{0} (u r) \\ r j_{2} (u r) \earr \right)
r^{3} \exp (- w r^{2}) = \frac{\sqrt{\pi}}{16 w^{7/2}}
\exp (- \frac{u^{2}}{4 w})
\left( \barr{c} 2 u w \\ 6 w - u^{2} \\ u^{2} \earr \right)
\eeqn

The only free parameter in the model is the
overall normalization of decays subsumed in $\kappa$  and the the combination
$\frac{a \tilde{c}}{9 \sqrt{3}} \frac{1}{2} A^{0}_{00} \sqrt{\frac{fb}{\pi}}$
in eqn. \ref{flux19731}.
However, if one repeats the analysis of \S \ref{section3}
but with the hybrid decay amplitude (eqn. \ref{hybriddec}) replaced by
that appropriate to ordinary mesons (eqn. \ref{mesondec}), one finds
that the same dimensionless combination controls the (known) decays
of the conventional mesons. A best fit gave a value of
 $0.64$ \cite{kokoski87} and we adopt this accordingly.
The scale of hybrid decays relative to ordinary meson decays are
then determined by $\kappa$: however, in the simplified framework
of ref. \cite{perantonis87} the estimated values for $N = 3-5$ beads
are $f=1.1$, $\kappa=0.9$ and $A^{0}_{00}=1.0$.

Our analytical calculation (with simplified wave functions)
reproduces an earlier numerical computation
 for light hybrids with exotic $J^{PC}$ \cite{kokoski85} to within
15\% on average. If we use the same hadron
masses as ref. \cite{kokoski85}, follow their
prescription (as outlined in ref. \cite{kokoski87}) of ignoring
all quark flavour symmetry breaking and
normalizing the decays as above, we find that the optimal comparison
with ref. \cite{kokoski85} follows with $\beta_{A}=0.27$ GeV
and $\betat=0.28$ GeV
throughout: this gives the widths in Table \ref{flux28470}.
We confirm their result that the decays indicated are dominant, except for
the case $J^{PC} = 0^{+-}$ where we find also prominent
 decays $ ( \pi , \omega ) 0^{+-} \rightarrow \kl K $ (with width
400 MeV) and $\pi 1^{-+} \rightarrow \kh K$
(with width 100 MeV) which were not listed in ref. \cite{kokoski85}.

Our analysis provides an
independent check on the results of ref. \cite{kokoski85}
and enables us to examine their sensitivity to the parameters. This merits
attention since the best fit to the widths of conventional mesons
by \cite{kokoski87} used a rather different value for $\beta$, namely
$\beta_{A} = \betat = 0.4$.
Indeed, this is in line with the modern preferred values
from harmonic oscillator basis approximations to meson spectroscopy e.g.
in the ISGW work \cite{isgur89}. Our preferred choice today is to adopt
the  harmonic oscillator basis fit  to
spin-averaged meson spectroscopy of ref. \cite{isgur89}.
Wherever values for $\beta$ are not available,
 we abstract them from the mean meson radii of Merlin \cite{merlin86}. We take
the string tension $b=0.18 \: {GeV}^{2}$, and the constituent-quark masses
$m_{u} = m_{d} = 0.33$ GeV, $m_{s}=0.55$ GeV and
 $m_{c}=1.82$ GeV.
Meson masses are taken from ref. \cite{pdg94}, and where
not available (as in the case of $^{3}P_{1} / \, ^{1}P_{1}$
mixing angles) we abstract them from spectroscopy predictions \cite{isgur85}
suitably adjusted relative to known masses.
Hybrid $\beta$'s, masses before spin splitting and
hyperfine splittings derive from Merlin
\cite{merlin86,merlin87}. Our quoted widths are computed for
$\delta = 0.62$ (though as mentioned earlier, the results with $\delta = 1$
are essentially similar to these).

We are able here, for the first time, to present also the most
prominant predicted widths
for both exotic and non-exotic $J^{PC}$ combinations. These are displayed
for $u,d,s$ flavours in tables \ref{flux71952} - \ref{flux29462} together
with the values assumed for
parameters.
One can
choose alternative values for these parameters and modify the widths
accordingly by use of table \ref{flux75423} and eqns. \ref{flux19731}
- \ref{flux82643}.

Application to hybrid charmonium decays
$c\bar{c} \: hybrid \rightarrow \dss D$ follows rather directly.
Their masses are predicted in the flux-tube
model to be $\approx 4.3$ GeV \cite{swanson94} which is in the vicinity of
 the $\dss D$ threshold. It is possible therefore that hybrid charmonium
will be kinematically forbidden from decaying into the preferred $(L=0)$ +
$(L=1)$ ($\{ D$ or $D^* \} + D^{**}$)
states, in which case their widths may be narrow and their signals
enhanced through decays into $\psi , \psi^{'} \cdots$ \cite{closecdf}.
Studies of hybrid charmonia will be reported
elsewhere \cite{page3}.


\section{Hybrid meson decay into two $L=0$  mesons \label{section5}}

For decays of hybrid mesons into two $L=0$ $q\bar{q}$ mesons,
the flux-tube model
predictions are very distinctive. When $\beta_{B} = \beta_{C}$ the
hybrid decay width is zero
 because the one unit of angular momentum of the incoming
hybrid around the $Q\bar{Q}$-axis cannot be absorbed by the angular momenta
of the outgoing mesons.
Non-zero widths arise if the S-wave hadrons have different size (a
result originally noted in the $^3P_0$ limit in ref. \cite{pene}).

Inserting eqn. \ref{hybriddec} into the master equation \ref{flux28943} and
performing the y-integration, only terms linear in
$\by$ in eqn. \ref{flux81639} contribute

\beqn
\int d^{3} \by \gamma (\br_{A},\by_{\perp}) \times (eqn. \, \ref{flux81639})
 =  - \frac{i \Delta}{\sqrt{2}} {\cal N}_{S}
{\cal N}_{B} {\cal N}_{C} \bar{\gamma}^{1}
\exp (- \frac{\tilde{\beta}^{2} \br_{A}^{2}}{4} + {(\frac{\br
\Delta}{4 \betat})}^{2} ) \:
{\cal D}^{1 \: \ast}_{\sigma \Lambda} (\phi,\theta,-\phi)
\eeqn
where $\bar{\gamma}^{1}$ is defined in eqn. \ref{flux20416}, and the
y-integration is done in the body-fixed system, introducing an
extra ${\cal D}$-function as in section 3. Clearly the decay amplitude
is proportional to $\Delta \equiv \beta_{B}^{2} - \beta_{C}^{2}$: when
 $\beta_{B} = \beta_{C}$, decay is prohibited.
Nonetheless, it is instructive to present the results scaled by the
factor $\Delta$. As we shall see, some of the widths would be
substantial were it not for this factor and hence it will be important
to consider the implications of a small, non-zero, value for $\Delta$
in hybrid meson phenomenology.

We  perform the r-integration in eqn. \ref{flux28943} as in
section 4. The partial wave amplitudes $M_{L} (A \rightarrow BC)$
are (modulo the factor ${(2 \pi)}^{3} \delta^{3} (\bp_{B} + \bp_{C})$)

\beqna
\label{flux19732}
\lefteqn{ M_{L} (A \rightarrow BC) =  -
(\frac{a \tilde{c}}{9 \sqrt{3}} \frac{1}{2} A^{0}_{00} \sqrt{\frac{fb}{\pi}})
\frac{\kappa \sqrt{b}}{{(1 + f b / (2 \betat^{2}) \, )}^{2}} } \\ \nonumber
& &  \eqntimes \Delta \, \sqrt{\frac{\pi}{3 \Gamma(3/2+\delta)}}
\frac{\beta_{A}^{3/2 + \delta} {(\beta_{B} \beta_{C})}^{3/2} }{ \betat^{5} }
\breve{M}_{L} (A \rightarrow B C)
\eeqna

In Table \ref{flux59263} we display the reduced partial wave amplitudes in
a compact form by defining

\beqn \begin{array}{lllll} \label{flux91042}
S = \breve{g}_{0} & P = -i \breve{g}_{1} & D = \breve{g}_{2}
                            &
F = 0 & G = 0
\end{array} \eeqn

\beqn
\breve{g}_{n} = \int_{0}^{\infty} dr
r^{2+\delta} j_{n}(\frac{M}{m+M} p_{B}r)
\exp (- ( 2 \beta_{A}^{2} + \betat^{2} -
{(\frac{\Delta}{2 \betat})}^{2} \, ) \frac{r^{2}}{4})
\eeqn

In Tables \ref{flux19302}-\ref{flux19304} we display a selection of
the most prominent widths calculated
from $M_{L} (A \rightarrow BC)$ in eqn. \ref{flux19732}, and scaled by the
dimensionless ratio
\beqn
\label{flux82053}
{(\frac{\Delta}{2 \betat ^2})}^{2} = {(\frac{\beta_{B}^{2}-\beta_{C}^{2}}
{\beta_{B}^{2}+\beta_{C}^{2}})}^{2}
\eeqn
We define the {\it intrinsic} width $\Gamma_{R}$ by $\Gamma_{R} (A
\rightarrow BC) \times (eqn. \: \ref{flux82053}) =
\Gamma (A \rightarrow BC)$.

In all cases the
same parameter values as in the corresponding flavour modes
in \S\ref{flux92712} and tables \ref{flux71952} - \ref{flux29462} are used.
The $\beta$'s
of \S\ref{flux92712}
are the same within the same hyperfine multiplet and so
would cause all widths in this section to be zero.
However,  estimates of $\beta$'s differing in the same hyperfine multiplet
can be found in the literature and these will lead to a non-zero value
for $\Delta$ and hence non-zero widths.

It is clear from tables \ref{flux19302} - \ref{flux19304} that some of the
widths would be substantial
were $\Delta$ non-zero. In some potential decay channels we would
expect $\Delta \neq 0$, for example, the $\pi$ is anomalously light and
may be expected to have an effective $\beta$ that differs significantly
from that of the $\rho$. Indeed, the ``effective" $\beta$ in ref.
\cite[Table I]{kokoski87} are used in tables \ref{flux19302} -
\ref{flux19304} to determine the widths $\Gamma$, and give

\beqn
\label{delta}
{(\frac{\Delta}{2 \betat ^2})}^{2} =0.2 \, (\pi \rho); \;  0.14 \,
(KK^*); \; 0.04 \, (DD^*)
\eeqn
Similar results obtain in the MIT Bag model \cite[fit II]{sharpe83}
by assuming $1 / \beta \propto$ {\em the
bag radius}.
In this context, note that the
{\it intrinsic} widths $\Gamma_{R} (A \rightarrow BC)$ are often predicted
to be substantial, e.g. for decays into $\pi + \rho; \; \pi + \omega$
and $K K^*$.
Indeed, values of $\sim 30\% $ larger arise
for $\rho \pi$ and $\omega \pi$ if one takes an alternative assumption
within the MIT Bag dynamics where  $\beta \propto m^{-1/3}$
(for massless quarks), but even with
the more conservative assumptions of eqn. \ref{delta} we see that we anticipate
significant couplings of hybrids in several of the ``forbidden" channels.

It is possible
therefore that hybrids could give rather distinctive signatures in
diffractive photoproduction or $e^+e^-$ annihilation, namely
the production of vector mesons in
$\pi\rho$ , $ \pi\omega$, $ KK^*$ or even $DD^*$ channels, but
absent (apart from mixing with conventional quarkonia) in the corresponding
$\pi\pi$, $\rho \rho$,$ K\bar{K}$, $D\bar{D}$ etc. final states.


\section{Phenomenology and Conclusions}

\vskip 0.2in

{\underline{$1^{-+}$}}

\vskip 0.1in

The most obvious signature for a hybrid meson is the appearance
of a flavoured state with an exotic combination for $J^{PC}$. Ref. \cite{lee94}
may have indications for such a state with $J^{PC} = 1^{-+}$ whose mass and
decay characteristics are in line with historical expectations.
The search was motivated by the selection rule (section 3) and
concentrated on the classic decay channel for $S+P$, namely
$\pi + f_1$, which is where the candidate has been sighted. The experiment sees
a broad structure in the mass region $1.6 - 2.2$ GeV which is suggestive
of being a composite of two objects at $1.7$ and $2.0$ GeV. It is the
latter that appears to have a resonant phase though they admit that
more data is required for a firm conclusion.

Our expectations for widths from
tables \ref{flux71952} and \ref{flux19302} for $J^{PC} = 1^{-+}$
at a mass of 2.0 GeV (which is essentially as originally predicted) are (in
MeV)
\beqn
\label{bnlwidth}
\pi f_1 : \pi b_1 : \pi \rho : \eta \pi : \eta' \pi \;
= \; 60 : 170 : 5-20 : 0-10 : 0-10
\eeqn
The former pair are similar to those in ref. \cite{kokoski85} but we note
also the possible presence of $\pi \rho$ or even $\pi \eta$
decays that are not negligible relative to the signal channel
$\pi f_1$. This may be important in view of a
puzzle, commented upon in ref. \cite{lee94}, that the production
mechanism appeared not to be as expected given the anticipated
hybrid dynamics. Instead of $b_1 $-exchange, leading to
the classic $S+P$ $\; \;  \pi + \bo$ coupling, significant $\pi + \rho$
coupling may be responsible. In view of our analysis in \S 5, and
eqn. \ref{bnlwidth} above,
it is clear that the latter coupling may be
significant on the scale of the $\pi f_1$ signal;
 the final state decays into $\pi +  \rho$ should
therefore also be investigated experimentally.
\vskip 0.2in
\noindent {\underline{$0^{-+} $}}

\vskip 0.1in

If this {\it prima facie} signal is indeed a resonant $1^{-+}$
hybrid excitation then one expects partners,
in particular $0^{-+}$, to be in this mass region. The VES Collaboration
sees an enigmatic
and clear $0^{-+}$ signal in diffractive production with 37 GeV
incident pions on beryllium \cite{ves}. They study the channels
$\pi^- N \rightarrow \pi^- \pi^+ \pi^- N; \; \pi^- K^+ K^- N$
and see a resonant signal $M \approx 1790$ MeV, $\Gamma \approx 200$ MeV
in the classic $(L=0)$ + $(L=1)$ $\bar{Q}q$
 channels $\pi^- + f_0; \; K^- + K^*_0, \; K {( K \pi )}_S $ with no
corresponding strong signal in the allowed $L=0$ two body
channels $\pi + \rho; \; K + K^*$.
The width and large couplings to kaons are both surprising if this
were the second radial excitation of the pion (the first radial
excitation is seen as a broad enhancement in accord with expectations).
Furthermore, the apparent preference for decay into $(L=0) + (L=1)$
mesons at the expense of $L=0$ pairs is qualitatively in
accord with expectations for hybrids.

Our quantitative estimates on the relative importance of available
channels further suppport this identification. For a $0^{-+}$ hybrid
at 1.8 GeV we find widths $\pi f_0(1300) \sim 170$ MeV; $\pi f_2 = 5-10$ MeV.
The $KK^*_0$ channel which is predicted to dominate for a 2.0 GeV initial
state (table \ref{flux71952})
 is kinematically suppressed though probably non-zero
due to the $\sim 300$ MeV width of the $K^*_0(1430)$. The
decay to $L=0$ pairs, which is naively expected to be suppressed,
turns out to be potentially significant,  $\pi \rho \sim 30$ MeV for a
1.8 GeV
$0^{-+}$ hybrid. This is compatible with the experimental limit
\beqn
\frac{0^{-+} \rightarrow \pi^- \rho^0}{0^{-+} \rightarrow\pi^- f_0(1300)}
\;  < \; 0.3 \; (95 \% \: C.L.)
\eeqn
The $KK^*(890)$ channel, by contrast, is expected to be
a mere $\sim 5$ MeV, which is consistent with the observed order of magnitude
suppression observed in ref. \cite{ves}
\beqn
\frac{0^{-+} \rightarrow K^- K^*}{0^{-+} \rightarrow (K^-K^+ \pi)_S} \;
 < \; 0.1 \; (95 \% \: C.L.)
\eeqn

The $\Gamma_{total} \sim 200 - 350$ MeV is also consistent with the
observed $200 \pm 50$ MeV. However, this may be fortuitous. First, the overall
scale of widths for hybrids, controlled by the breaking of the excited
flux-tube, may differ from that of the ground state conventional decays such
that all hybrid predictions will need to be rescaled by an overall
constant. Furthermore our calculations are all in
the narrow width approximation whereas the $f_0(1300)$ at least is a
broad ill defined structure.
The precise role of the enigmatic $f_0(980)$, to which the resonance
also appears to couple experimentally, also perturbs a detailed analysis
at this stage. The data here are
\beqn
\frac{0^{-+} \rightarrow \pi^- f_0(980)}{0^{-+} \rightarrow \pi^- f_0(1300)}
\;  = \; 0.9 \pm 0.1
\eeqn
As noted in ref. \cite{ves} this is an unexpectedly high value since the
$f_0(980)$ has a small width and strong coupling to strangeness while the
$f_0(1300)$ is a broad object coupled mainly to non-strange quarks.
However, this may be natural for a hybrid at this mass for the following
reason.
The strongest predicted decay path  (see table \ref{flux71952}) would be
$0^{-+} \rightarrow KK^*_0$ but for the fact that this is below threshold
for the 1.8 GeV initial state, thus the $(KK\pi)_S$ is expected to be
significant (as observed \cite{ves}) and, at some level, may feed the channel
$\pi f_0(980)$ through the strong affinity of $K\bar{K} \rightarrow f_0(980)$.
Thus the overall expectations are in line with
the data. Important tests are now that there should be
a measureable coupling to the $\pi \rho$ channel with only a small
$\pi f_2$ or $KK^*$ contribution.
\vskip 0.2in
\noindent {\underline{$2^{-+} $}}

\vskip 0.1in

This suppression of $\pi f_2$ for the $0^{-+}$ is quite opposite to
the prediction for the $2^{-+}$ partner for which this channel should
dominate significantly over the $\pi f_0$ partner (not least because of the
interchanged role of $S$ and $D$ waves). This is a problem if one wishes
to identify the $2^{-+}$ seen at $\sim 2.2$ GeV
at VES as the hybrid partner of the $0^{-+}$. The putative
signal is claimed in $\pi f_0(1300) $ whereas
no $\pi f_2$ nor $\pi f_0(980)$ are reported.
The properties and existence of this state are less clearcut experimentally and
on mass alone it could qualify either as a radial excitation or
tantalisingly in accord with the emergence of a family of hybrids.
However, as alluded to above,
 its decay channels do not appear superficially to be in line with
those expected for hybrids.  The $\pi f_0$ is predicted to be small
while that to $\pi f_2$, $\pi b_1$ together with $KK^*$ or $\pi a_2$
provide the
anticipated signals. From the regularities in table 1 we see immediately
the source of the pattern
\beqn
\pi f_0 :\pi f_2 \; = \; 1 : 7
\eeqn
for the D-waves, let alone the S-wave contribution for $\pi f_2$.
If $\pi f_0(1300) \geq \pi f_2(1270)$ is sustained for this state,
it is either not a hybrid or there is some new dynamics connecting
it to the broad $f_0(1300)$ state.

Historically the ACCMOR Collaboration
\cite{accmor} has argued for a $2^{-+}$ state around 2.1 GeV, or
possibly 1.8 GeV, coupled to $\pi f_2$, which was used to set the mass
scale in a Bag Model simulation of hybrids in ref. \cite{sharpe83}.
The lower mass is tantalisingly
similar to sightings of a possible $2^{-+}$ in photoproduction via
$\pi$ exchange \cite{condo} and coupled to $\pi \rho$ and $\pi f_2$.
These suggest that there may be interesting activity in the $2^{-+}$
wave which may herald new degrees of freedom; if hybrid components are
present in this (these?) state, we urge a search for the $\pi b_1$
decay channel which, at the lower mass, could have a branching ratio of
up to $50 \%$.
\vskip 0.2in
\noindent {\underline{$1^{--} $}}

\vskip 0.1in

If these are indeed signalling the emergence of the lowest lying
families of
hybrids, then there must be a nonet of $1^{--}$ partners. As the $0^{-+}$
appeared in diffractive $\pi$ production, so we anticipate the
appearance of the $1^{--}$ in diffractive photoproduction or
$e^+e^-$ annihilation. We advocate searching for the lightest vector
hybrids in
\beqn
\gamma (p) \rightarrow \pi a_1 (p) \rightarrow 4\pi (p)
\eeqn
where within our harmonic oscillator approximation we predict for an
isovector (in MeV)
\beqn
\pi a_1 : \pi a_2 : \pi h_1 : \rho \rho : \pi \omega : \pi \pi \; =
\; 170 : 50 : 0 : 0 : 10-20 : 0
\eeqn
Alternatively, for a vector hybrid at a mass of $\sim 1.5$ GeV (see
below) these become
\beqn
\pi a_1 : \pi a_2 : \pi h_1 : \rho \rho : \pi \omega : \pi \pi \; =
\; 140 :\, \, \sim 0 : 0 : 0 : 5-10 : 0
\eeqn
These are very different from the predictions of radial or $^3D_1$
decays of quarkonia \cite{kokoski87,isgur85,busetto}.
In particular
the suppression of $\pi h_1$ relative to $\pi a_1$ is, within the flux-tube
model, a crucial test of the hybrid initial state in contrast to the case
of a $^3D_1$ or radially excited $1^{--}$  for which the $\pi h_1$
would be expected to dominate over $\pi a_1$ \cite{kokoski87,busetto}.
The reason is that in the hybrid $1^{--}$
the $Q\bar{Q}$ have $S=0$, whereas for the ``conventional
quarkonium" $1^{--}$ the $Q\bar{Q}$ have $S=1$; the $^3P_0$ decay is forbidden
by spin orthogonality in the former example for final states where the
mesons' $Q\bar{Q}$ have $S=0$, as in the $\pi h_1$ example.
 It is therefore
interesting that the detailed analyses of refs. \cite{don2,don1} comment
on the apparently anomalous decays that they find for
the $1^{--}$ state $``\rho_1"(1450)$, in particular the suppression of $\pi
h_1$
relative to $ \pi a_1$ and the dominance of the latter over the $\pi \omega$ :
\beqn
\pi a_1 : \pi h_1  +  \rho \rho : \pi \omega : \pi \pi \;  =
\; 190 : 0-39 : 50-80 : 17 -25
\eeqn
 It is noticeable
that the $\pi \pi$ decay also is strongly suppressed though non-zero; if this
is substantiated it could indicate either a deviation from the harmonic
wave function approximation or in addition some mixing between hybrid and
radial
vector mesons in this region. The latter could also rather naturally
explain the enhancement of the $\pi \omega$ channel
as well as the repulsion of the
eigenstate to low mass. This is beyond the present work
but merits further attention in view of the fact that the decay channels
of the $\rho_1$, in particular the large $\pi a_1$ component and suppressed
$\pi \pi$,
require that ``mixing with non $Q\bar{Q}$ states must
occur" \cite{don2}\footnote{It is interesting that there appear to be
possible solutions to the data with $\pi \pi$ even more suppressed and
the $\pi a_1$ increased in compensation (A. Donnachie, private communication)}.
We suggest that a detailed comparison of $e^+e^-$ with diffractive
photoproduction may help to isolate the hybrid contributions more
clearly as the relative abundance of hybrid excitation and quarkonium
production is in general expected to differ in the two cases: as
diffractive photoproduction involves the transition $\gamma \rightarrow
Q\bar{Q}$ in the probable presence of a gluonic Pomeron, there is the
possibility of ``plucking the string".

Ref. \cite{don2} also finds evidence for $\omega (1440)$ with no visible
decays into $\pi b_1$ which is in significant contrast to the expectations
for conventional $Q\bar{Q} \; (^3S_1$ or $^3D_1)$ initial states. In the hybrid
interpretation this suppression is natural and
is the isoscalar analogue of the $\pi h_1$ selection rule
alluded to above. It is also interesting to note that for a hybrid
$\omega (1440)$, the ``wannabee" $(L=0) + (L=1)$ decay paths are kinematically
suppressed leaving the $\pi \rho$ and possibly $\eta \omega$ as dominant
decays.

Insofar as $L=0$ pairs are predicted to be
suppressed but not totally absent in the decay
products,
searches for $\pi \rho$; $ KK^*$; $ \pi \omega \: (\pi \eta)$ should be
made. Confirmation of signals in such channels together with
them being dominated by $(L=0) + (L=1)$ states would add considerable
weight to the hybrid hypothesis. We need more detailed study of decays of
radial
excitations in the quark model to see if they imitate the
hybrid preferences for $S+P$ modes: as noted above for the $1^{--}$
channel, the relative branching ratios to these can be distinctive as in the
case of $\pi h_1 : \pi a_1$ which differ appreciably for $\rho_{hybrid}$ and
$\rho_{conventional}$.  If these are hybrid states then
necessarily there will be partners whose production and decay channels
become rather tightly constrained.

To the extent that signals are appearing in the expected mass region
for light flavours, together with hints of a rich $0^{++}$
spectroscopy in the mass region anticipated for
gluonic excitations in the pure gauge sector, we have increasing confidence
in predictions for the gluonic excitations in more generality, in
particular for hybrids containing
heavy flavours e.g. $c\bar{c}$ and $b\bar{b}$. These are predicted to
occur in the vicinity of charm threshold \cite{swanson94,perantonis90} and so
we
advocate intensive study of this region, in particular with $e^+e^-$.
Rather clear signals and
the clean environment may distinguish radial from hybrid here.
The $S+S$ suppression is more dramatic than for light flavours
and so there is the exciting possibility that
hybrid charmonium will be narrow ($\sim 1 - 10$ MeV). Appearance
of states above charm threshold decaying into $DD^*$ but strongly suppressed
or even absent in
$D\bar{D}$, $D^*\bar{D^*}$ would be rather striking.

\section{Acknowledgements}
We thank T. Barnes, S.U. Chung, A. Donnachie, and J. Paton
for discussions and comments.
This work has been supported in part by the European Community Human
Mobility Program ``Eurodafne", Contract CHRX-CT92-0026 and (PP) by a
scholarship from the University of Cape Town.


\section{Appendix A : General decay formalism}

Consider a quark-antiquark bound system A with quark at position
$\br$ and antiquark
at $\bar{\br}$ with masses $m$ and $\bar{m}$ respectively. The system has a
momentum $\bp_{A}$ and wavefunction $\psi_{A}^{LM_{L}
\Lambda} (\bar{\br}-\br)$, with
relative coordinate $\br_{A} \equiv \bar{\br}-\br$,
angular momentum $L_{A}$ and projection $M_{L}^{A}$. The flux-tube has
$\Lambda$
units of angular momentum around the q\={q}-axis.
Introduce a
second-quantized formalism in which the normalized
wavefunction is written in the
L-S basis as

\beqna
\label{flux83624}
\lefteqn{\mid \! A \rangle \equiv \: \mid \! A^{C \, F \, SM_{S} \,
LM_{L} \Lambda}
_{ \{ n_{m+},n_{m-} \} } (\bp_{A}) \rangle =
\sum_{f \bar{f} s \bar{s} c \bar{c}}
\int d^{3} \br \, d^{3} \bar{\br} \, A_{c \bar{c}}^{C} A_{f \bar{f}}^{F}
A_{s \bar{s}}^{SM_{S}} \psi_{A}^{LM_{L} \Lambda} (\br_{A})}
\nonumber \\
& & \mbox{} \times \exp (i \bp_{A} \cdot \frac{m \br + \bar{m}
\bar{\br}}{m + \bar{m}}) \:
q^{+}_{c f s} (\br) \: \bar{q}^{\, +}_{\bar{c} \bar{f} \bar{s}} (\bar{\br})
\mid \! 0 \rangle
\: \otimes \: \mid \! \br_{A} \, \{ n_{m+},n_{m-} \} \rangle
\eeqna

with $A^{C}$, $A^{F}$ and $A^{SM_{S}}$ referring to the colour,
flavour and spin matrices
respectively. Here $q^{+}_{c f s} (\br)$ and
$\bar{q}^{\, +}_{c f s} (\br)$ are the non-relativistic position space
quark and antiquark creation operators respectively, obeying anticommutation
relations of the type $\{ q^{+}_{c f s} (\br) ,
q^{+}_{\bar{c} \bar{f} \bar{s}} (\bar{\br}) \} = \delta^{3}(\br - \bar{\br})
\, \delta_{c \bar{c}} \delta_{f \bar{f}} \delta_{s \bar{s}}$.
The state $\mid \! \br_{A} \, \{ n_{m+},n_{m-} \} \rangle$
represents the system of $N$
beads a distance $a$ apart in modes $\{ n_{m+},n_{m-} \}$
\cite{paton85} ,
vibrating w.r.t.\ the q\={q} - axis $\br_{A}$ as equilibrium position.
The beads are connected to each other and the quarks at the ends via
a non-relativistic string potential with string tension $b$.

The $^{3}P_{0}$ quark-antiquark creation operator $\hat{\cal C}$
motivated from the strong coupling expansion of
Hamiltonian lattice gauge theory \cite{kokoski87} is

\beqna
\label{flux83625}
\hat{\cal C} &  = & \frac{a \tilde{c}}{9} \sum_{cfs \bar{s}} \int d^{3} \bx \,
b(\bx) \psi^{+}_{cfs}(\bx)
\balpha_{s \bar{s}} \cdot \bnabla \psi_{cf \bar{s}}(\bx) \nonumber \\
& = & \frac{a \tilde{c}}{9} \sum_{cfs \bar{s}} \int d^{3} \bx \,
b(\bx) q^{+}_{cfs}(\bx)
\bsigma_{s \bar{s}} \cdot \bnabla \bar{q}^{\, +}_{cf \bar{s}}(\bx)
\eeqna

where we restrict $\hat{\cal C}$ to q\={q}-creation in the last line.
Here $\psi_{cfs}(\bx)$ is the usual relativistic Dirac fermion operator
with $\balpha$ the Dirac matrices defined as usual in terms of \ the Pauli
matrices $\bsigma$.
A bead is annihilated by $b(\bx)$ at the pair creation position.
The identity can be established by defining $q^{+}_{cfs}(\bx)$
i.t.o.\ the quark creation operators of the momentum space expansion of
$\psi_{cfs}(\bx)$. The factor of $1/9$ arises by requiring the annihilated
flux to couple to a singlet and be unoriented \cite{kokoski87}. We
introduce an unknown constant $\tilde{c}$ and lattice spacing $a$, so that
$\dim (\hat{\cal C}) = 1$, as required.

We can now rigorously define the flux-tube overlap $\gamma
(\br_{A},\by_{\perp}) \equiv
\langle \br_{A} \, \{ n_{m+},n_{m-} \} \! \mid b(\by)
\mid  \! \br_{B} \br_{C} \, \{ n_{m+},n_{m-} \} \rangle$ introduced in
\S\ref{section2}.

\begin{table}[p] \begin{center}
\caption{Partial wave amplitudes $\tilde{M}_{L} (A \rightarrow B C)$
written in terms of the functions defined in eqn. \protect\ref{flux84610}
and named in accordance with partial wave S, P, D, F or G.
We display various $J^{PC}$ of
the initial hybrid A decaying into a L=1 meson B and pseudoscalar meson C.
Starred amplitudes vanish even with non-S.H.O. radial wave functions.}
\label{flux75423}
\begin{tabular}{|l|l|c||l|l|c||l|l|c|}
\hline 
A & B & $\tilde{M}_{L}$ &
A & B & $\tilde{M}_{L}$ &
A & B & $\tilde{M}_{L}$ \\
\hline \hline 
$2^{-+}$ & $2^{++}$ &       $ - \sqrt{5} S / \sqrt{18}$   &
$1^{-+}$ & $2^{++}$ &       $ 0 \times D       $   &
$1^{+-}$ & $2^{++}$    &    $ - P_{3} / \sqrt{15}$ \\
$      $ & $      $ &       $ - \sqrt{7} D / 3        $   & $      $ &
$1^{++}$ &       $ S / \sqrt{6}     $   & $      $ & $      $    &
    $ - F / \sqrt{10}$ \\
$      $ & $      $ &       $          G              $   & $      $ &
$      $ &       $ -D / \sqrt{3}    $   & $      $ & $1^{++}$    &    $
P_{2}$ \\
$      $ & $1^{++}$ &       $ 0 \times D              $   & $      $ &
$1^{+-}$ &       $ S / \sqrt{3}     $   & $      $ & $0^{++}$    &
$ P_{1} / \sqrt{3}$ \\
$      $ & $0^{++}$ &       $ D / 3                   $   & $      $ &
$      $ &       $ D / \sqrt{6}     $   & $      $ & $1^{+-}$    &
$ - P_{1} / \sqrt{2}$ \\
\cline{4-9}
$      $ & $1^{+-}$ &       $ - D / \sqrt{2}          $   & $0^{-+}$
& $2^{++}$ &       $ D / 3            $   & $0^{+-}$ & $1^{++}$    &
$ - P_{1} / \sqrt{3}$ \\
\cline{1-3}
$2^{+-}$ & $2^{++}$ &       $ P_{5} / \sqrt{5}        $   & $      $ &
$0^{++}$ &       $ \sqrt{2} S / 3   $   & $      $ & $1^{+-}$    &
$ \sqrt{2} P_{2} / \sqrt{3}$ \\
\cline{4-9}
$      $ & $      $ &       $ -F / \sqrt{5}           $   & $1^{--}$
& $2^{++}$ &       $ D / \sqrt{2}     $   & $1^{++}$ & $2^{++}$    &
    $ - P_{4} / \sqrt{30}$ \\
$      $ & $1^{++}$ &       $ P_{3} / \sqrt{15}       $   & $
      $ & $1^{++}$ &       $ S / \sqrt{3}     $   & $      $ & $      $
    &    $ F / \sqrt{5}$ \\
$      $ & $      $ &       $ F / \sqrt{10}           $   & $      $ &
$      $ &       $ D / \sqrt{6}     $   & $      $ & $1^{++}$    &
$ - P_{1} / \sqrt{2}$ \\
$      $ & $1^{+-}$ &       $ P_{4} / \sqrt{30}       $   & $      $ &
$1^{+-}$ &       $ 0\times S ^{\ast}$   & $      $ & $0^{++}$   &
$ -\sqrt{2} P_{2} / \sqrt{3}$ \\
$      $ & $      $ &       $ - F / \sqrt{5}          $   & $      $ & $
      $ &       $ 0\times D ^{\ast}$   & $      $ & $1^{+-}$    &
$ 0 \times P ^{\ast}$ \\
\hline 
\end{tabular}
\end{center} \end{table}

\btablec
\caption{Widths in MeV for $hybrid \: A \rightarrow B C$ for exotic hybrid
$J^{PC}$ in partial wave $L$. Here $\pi$, $\omega$ and $\phi$ indicate
flavour states $\protect\sqrt{\frac{1}{2}}(u\bar{u}-d\bar{d})$,
$\protect\sqrt{\frac{1}{2}}(u\bar{u}+d\bar{d})$ and $s\bar{s}$ respectively.
We adopted hybrid masses of 1.9 GeV ($\pi , \omega$) and 2.1 GeV ($\phi$);
a $^{3}P_{1} / \: ^{1}P_{1}$ mixing of $45^{o}$ in the P-wave kaon
sector; and assumed $f=1$, $\kappa = 1$, $\delta = 1$ in order to
{\it compare} with the
widths $\Gamma_{2}$ of ref. \protect\cite{kokoski85}. Our optimal fit to ref.
 \protect\cite{kokoski85}
gives widths $\Gamma_{1}$ (see \S \protect\ref{flux92712}). }
\label{flux28470}
\begin{tabular}{|l|l|c|r|r||l|l|c|r|r|}
\hline 
$A$ & $B,C$ & $L$ & $\Gamma_{1}$ & $\Gamma_{2}$ &
$A$ & $B,C$ & $L$ & $\Gamma_{1}$ & $\Gamma_{2}$ \\
\hline \hline 
 $\pi   1^{-+}$ &$\bo\pi$ & S &    100 & 100 & $\phi   1^{-+}$ & $\kl K $
& D &   90 & 80\\
                &         & D &     20 &  30 &
& $\kh K $ & S &  200 & 250\\ \cline{6-10}
                &$\fo\pi$ & S &     30 &  30
& $\pi    0^{+-}$ & $\ao\pi$ & P &  600 & 800\\ 
                &         & D &     20 &  20
&                 & $\ho\pi$ & P &  100 & 100\\ \cline{1-10}
$\omega 1^{-+}$ &$\ao\pi$ & S &     90 & 100
& $\omega 0^{+-}$ & $\bo\pi$ & P &  250 & 250\\ \cline{6-10}
                &         & D &     60 &  70
& $\phi   0^{+-}$ & $\kl K $ & P &  500 & 800 \\
                &$\kh K $ & S &    100 & 100
&                 & $\kh K $ & P &   70 & 50\\ \cline{1-10}
$\pi    2^{+-}$ &$\at\pi$ & P &    350 & 450
& $\omega 2^{+-}$ & $\bo\pi$ & P &  350 & 500\\ \cline{6-10}
                &$\ao\pi$ & P &    100 & 100
& $\phi   2^{+-}$ & $\kt K $ & P &  300 & 250\\
                &$\ho\pi$ & P &    125 & 150
&                 & $\kh K $ & P &  250 & 200\\
\hline 
\end{tabular}
\etablec

\btablec
\caption{Dominant widths in MeV for
$\protect\sqrt{\frac{1}{2}}(u\bar{u}-d\bar{d}) \: hybrid \: A
\rightarrow B C$ for various
$J^{PC}$ in partial wave $L$.
The quark model assignments for the
mesons are those of the
PDG tables \protect\cite{pdg94}. All $\beta$'s are rescaled from
the ISGW / Merlin
values by $5/4$ to form ``effective'' $\beta $'s consistent with that
of $\beta = 0.4$.
Hybrid masses before spin splitting are 2.0 GeV, except for $0^{+-}$
(2.3 GeV), $1^{+-}$ (2.15 GeV) and $2^{+-}$ (1.85 GeV), following
ref. \protect\cite{merlin87}.
Final states containing $\pi$ have $\protect\betat = 0.36$ GeV,
otherwise $\protect\betat = 0.40$ GeV. For the hybrid we use
$\beta_{A}$ = 0.27 GeV.
$\eta$ indicates $\protect\sqrt{\frac{1}{2}}(u\bar{u}+d\bar{d})$ at 550 MeV.
The $^{3}P_{1} / \: ^{1}P_{1}$-mixing is $34^{o}$ in the $L_{B}=1$
kaon sector. }
\label{flux71952}
\label{flux38713}
\label{flux55458}
\begin{tabular}{|l|l|c|r||l|l|c|r||l|l|c|r|}
\hline 
$A$ & $B,C$ & $L$ & $\Gamma$ &
$A$ & $B,C$ & $L$ & $\Gamma$ &
$A$ & $B,C$ & $L$ & $\Gamma$ \\
\hline \hline 
$2^{-+}$ & $ \ft \pi  $ & S &     40 &$1^{+-}$ & $ \at \pi  $
& P &    175 &$1^{-+}$ & $ \fo \pi  $ & S &     40 \\
         &              & D &     20 &         & $ \ao \pi  $
& P &     90 &         &              & D &     20 \\
         & $ \bo \pi  $ & D &     40 &         & $ \ho \pi
$ & P &    175 &         & $ \bo \pi  $ & S &    150 \\
         & $ \at \eta $ & S &$\sim 40$ &         & $ \bo \eta $ & P &
    150 &         &              & D &     20 \\
         & $ \kt K    $ & S &$\sim 30$ &         & $ \kt K    $ & P &
60 &         & $ \ao \eta $ & S &     50 \\ \cline{1-4}
$2^{+-}$ & $ \at \pi  $ & P &    200 &         & $ \kl K    $ & P
&    250 &         & $ \kl K    $ & S &     20 \\
         & $ \ao \pi  $ & P &     70 &         & $ \kz K    $
& P &     70 &         & $ \kh K    $ & S &$\sim 125$\\ \cline{5-12}
         & $ \ho \pi  $ & P &     90 &$1^{++}$ & $ \ft \pi
$ & P &    175 &$0^{-+}$ & $ \ft \pi  $ & D &     20 \\
         & $ \bo \eta $ & P &$\sim 15$ &
& $ \fo \pi  $ & P &
150 &         & $ \fz \pi  $ & S &$\sim 150$ \\ \cline{1-4}
$0^{+-}$ & $ \ao \pi  $ & P &    700 &         & $ \fz \pi  $
& P &$\sim 20$&         & $ \kz K    $ & S &$\sim 200$\\ \cline{9-12}
         & $ \ho \pi  $ & P &    125 &         & $ \at \eta $ & P
&     50 &$1^{--}$ & $ \at \pi  $ & D &     50 \\
         & $ \bo \eta $ & P &     80 &         & $ \ao \eta $
& P &     90 &         & $ \ao \pi  $ & S &    150 \\
         & $ \kl K    $ & P &    600 &         & $ \kt K    $
& P &$\sim 20$ &         &              & D &     20 \\
         & $ \kh K    $ & P &    150 &         & $ \kl K    $
& P &     40 &         & $ \kl K    $ & S &     40 \\
         &              &   &        &         & $ \kh K    $
& P &$\sim 20$ &         & $ \kh K    $ & S &$\sim 60$ \\
\hline 
\end{tabular}
\etablec

\btablec
\caption{As in table \protect\ref{flux71952} but for initial hybrid
$\protect\sqrt{\frac{1}{2}}(u\bar{u}+d\bar{d})$. }
\label{flux39156}
\label{flux32419}
\begin{tabular}{|l|l|c|r||l|l|c|r||l|l|c|r|}
\hline 
$A$ & $B,C$ & $L$ & $\Gamma$ &
$A$ & $B,C$ & $L$ & $\Gamma$ &
$A$ & $B,C$ & $L$ & $\Gamma$ \\
\hline \hline 
$2^{-+}$ & $ \at \pi  $ & S &    125 &$2^{+-}$ & $ \bo \pi  $
& P &    250 &$1^{++}$ & $ \at \pi  $ & P &    500 \\
         &              & D &     60 &         & $ \ho \eta $ & P &
30 &         & $ \ao \pi  $ & P &    450 \\ \cline{5-8}
         & $ \ft \eta $ & S &$\sim 50$ &$0^{+-}$ & $ \bo \pi
$ & P &    300 &         & $ \ft \eta $ & P &     70 \\
         & $ \kt K    $ & S &$\sim 30$ &         & $ \ho \eta $ & P &
90 &         & $ \fo \eta $ & P &     60 \\ \cline{1-4}
$1^{+-}$ & $ \bo \pi  $ & P &    500 &         & $ \kl K    $ & P
&    600 &         & $ \kt K    $ & P &$\sim 20$ \\
         & $ \ho \eta $ & P &    175 &         & $ \kh K    $ & P &150
&         & $ \kl K    $ & P &     40 \\ \cline{5-8}
         & $ \kt K    $ & P &     60 &$1^{-+}$ & $ \ao \pi  $ & S
&    100 &         & $ \kh K    $ & P &$\sim 20$ \\ \cline{9-12}
         & $ \kl K    $ & P &    250 &         &              & D
&     70 &$0^{-+}$ & $ \at \pi  $ & D &     60 \\
         & $ \kz K    $ & P &     70 &         & $ \fo \eta $ & S &50
&         & $ \fz \eta $ & S &$\sim 200$ \\ \cline{1-4}
$1^{--}$ & $ \kl K    $ & S &     40 &         & $ \kl K    $ & S
&     20 &         & $ \kz K    $ & S &$\sim 200$ \\
         & $ \kh K    $ & S &     60 &         & $ \kh K    $ & S
&$\sim 125$ & & & & \\
\hline 
\end{tabular}
\etablec

\btablec
\caption{As in table \protect\ref{flux71952} but for an initial
$s\bar{s}$-hybrid.
Hybrid masses before spin splitting are 2.15 GeV, except for $0^{+-}$
(2.25 GeV).
Final states containing $K$ have $\protect\betat = 0.40$ GeV,
otherwise $\protect\betat = 0.44$ GeV. For the hybrid we use $\beta_{A}$
= 0.30 GeV.}
\label{flux29462}
\label{flux41493}
\begin{tabular}{|l|l|c|r||l|l|c|r||l|l|c|r|}
\hline 
$s\bar{s}g$ & $B,C$ & $L$ & $\Gamma$ &
$s\bar{s}g$ & $B,C$ & $L$ & $\Gamma$ &
$s\bar{s}g$ & $B,C$ & $L$ & $\Gamma$ \\
\hline \hline 
$2^{-+}$ & $ \kt K     $ & S &    100 &$1^{-+}$ & $ \kl K     $ & S
&     40 &$0^{+-}$ & $ \kl  K    $ & P &     400\\
         & $ \kl  K    $ & D &     20 &         &               & D &
60  &         & $ \kh  K    $ & P &     175\\ \cline{1-4} \cline{9-12}
$1^{+-}$ & $ \kt  K    $ & P &      70&         & $ \kh K     $ & S
&    250 &$0^{-+}$ & $ \kt  K    $ & D &     20 \\ \cline{5-8}
         & $ \kl  K    $ & P &     250&$2^{+-}$ & $ \kt  K    $
& P &     90 &         & $ \kz  K    $ & S &    400 \\ \cline{9-12}
         & $ \kz  K    $ & P &     125&         & $ \kl  K    $ & P &
30 &$1^{--}$ & $ \kt  K    $ & D &     20 \\ \cline{1-4}
$1^{++}$ & $ \kt  K    $ & P &     125&         & $ \kh  K    $
& P &     70 &         & $ \kl  K    $ & S &     60 \\
         & $ \kl  K    $ & P &      70&         &
&   & &         & $ \kh   K   $ & S &    125 \\
         & $ \kh  K    $ & P &     100&         &               &   && & & & \\
\hline 
\end{tabular}
\etablec

\btablec
\caption{Partial wave amplitudes $\breve{M}_{L} (A \rightarrow B C)$
indicated in terms of the functions defined in eqn. \protect\ref{flux91042}
and named in accordance with partial waves S, P, D, F or G.
We display various $J^{PC}$ of the initial hybrid A decaying into
pseudoscalar $0^{-+}$ (P) or vector $1^{--}$ (V) final mesons.
Starred amplitudes vanish even with non-S.H.O. radial wave functions.}
\label{flux59263}
\begin{tabular}{|l|l|c||l|l|c||l|l|c|}
\hline 
A & BC & $\breve{M}_{L}$ &
A & BC & $\breve{M}_{L}$ &
A & BC & $\breve{M}_{L}$ \\
\hline \hline 
$2^{-+}$ & VP &       $ -\sqrt{15} P / \sqrt{2} $   & $1^{--}$ & PP
&       $ 0\times P ^{\ast}$   & $1^{+-}$ & VP  &    $ 2 \sqrt{3} S $ \\
	 &    &       $          F              $   &          & VP
&       $ 3 P              $   &          &     &
$ \sqrt{3} D / \sqrt{2} $ \\
	 & VV &       $ 3 \sqrt{5} P            $   &
& VV &       $ 3    \sqrt{2} P  $   &
& VV  &    $  0 \times S ^{\ast} $ \\
	 &    &       $          F              $   &          &
&       $ 0\times F ^{\ast}$   &          &     &    $ 3 D$ \\ \cline{1-9}
$1^{-+}$ & PP &       $     3 P                 $   & $2^{+-}$ & PP
&       $     \sqrt{3} D   $   & $0^{+-}$ & PP  &
$ - \sqrt{6} S $ \\
	 & VP &       $ 3 P / \sqrt{2}          $   &
& VP &       $ 3 D / \sqrt{2}   $   &          & VV  &
$ - \sqrt{2} S $ \\
	 & VV &       $ 3       \sqrt{2} P      $   &          & VV
&       $ 2 \sqrt{10} S    $   &          &     &
$ - 2 D               $  \\  \cline{7-9}
	 &    &       $  F                      $   &          &
&       $ 2   \sqrt{2} D   $   & $1^{++}$ & VP  &    $ \sqrt{6} S $
\\ \cline{1-3}
$0^{-+}$ & VP &       $ \sqrt{6 } P             $   &          &
&       $          G       $   &          &     &    $ - \sqrt{3} D $
\\
	 & VV &       $ 0 \times P ^{\ast}      $   &
&    &       $                  $   &          & VV  &
$ 2 \sqrt{3} S $                 \\
	 &    &                                     &          &
&                              &          &     &    $ -\sqrt{6} D $ \\
\hline 
\end{tabular}
\etablec

\btablec
\caption{Dominant widths in MeV for
$\protect\sqrt{\frac{1}{2}}(u\bar{u}-d\bar{d}) \: hybrid \: A
\rightarrow B C$, where B and C are both L=0 quarkonia.
$\Gamma = \Gamma_{R} \times (eqn. \: \protect\ref{flux82053})$.
$\eta (\eta^{'})$ indicates $\protect\sqrt{\frac{1}{2}}(u\bar{u}+d\bar{d})$
at 550 MeV
(960 MeV) respectively. Starred $\Gamma$'s tend to be $\leq 1$ MeV, and
are highly sensitive to model dependent assumptions about final state
$\beta $'s. This table is the corrected version of table 7 in
OUTP-94-29P.}
\label{flux19302}
\begin{tabular}{|l|l|c|r|r||l|l|c|r|r||l|l|c|r|r|}
\hline 
$A$ & $B,C$ & $L$ & $\Gamma_{R}$ & $\Gamma$ &
$A$ & $B,C$ & $L$ & $\Gamma_{R}$ & $\Gamma$ &
$A$ & $B,C$ & $L$ & $\Gamma_{R}$ & $\Gamma$ \\
\hline \hline 
$2^{-+}$ & $ \rho \pi     $ & P &40 &   8 &
$0^{-+}$ & $ \rho \pi    $ & P &150 &  30 &
$1^{--}$ & $ \omega \pi  $ & P & 40 &   8 \\
	 & $ K^{\ast} K   $ & P &15 &   2 &
         & $ K^{\ast} K  $ & P & 60 &   8 &
	 & $ \rho \eta    $ & P &30 &   7 \\ \cline{6-10}
	 & $ \rho \omega  $ & P & 70&$\ast$&
$1^{-+}$ &$ \eta \pi    $ & P & 40 &$\ast$&
	 & $ \rho \eta^{'}$ & P & 15 &   3 \\ \cline{1-5}
$1^{+-}$ & $ \omega \pi$   & S & 70 &  15&
         & $ \eta^{'} \pi$ & P & 40 &$\ast$&
         & $ K^{\ast} K  $ & P & 30 &   4 \\ \cline{11-15}
         & $ \rho \eta    $ & S &100&  20&
         & $ \rho \pi    $ & P & 40 &   8&
$1^{++}$ & $ \rho \pi     $ & S & 80&  20 \\
         & $ \rho \eta^{'}$ & S &150&  30&
         &$ K^{\ast} K  $  & P & 15 &   2&
	 & $ K^{\ast} K   $ & S &125&  15 \\
         & $ K^{\ast} K  $ & S & 200&  30&
         & $ \rho \omega $ & P & 50 &$\ast$ &
	 & $ \rho \omega  $ & S &125&$\ast$\\
\hline 
\end{tabular}
\etablec

\btablec
\caption{As in table \protect\ref{flux19302}
 but for initial hybrid
$\protect\sqrt{\frac{1}{2}}(u\bar{u}+d\bar{d})$.  }
\label{flux19303}
\begin{tabular}{|l|l|c|r|r||l|l|c|r|r||l|l|c|r|r|}
\hline 
$A$ & $B,C$ & $L$ & $\Gamma_{R}$ & $\Gamma$ &
$A$ & $B,C$ & $L$ & $\Gamma_{R}$ & $\Gamma$ &
$A$ & $B,C$ & $L$ & $\Gamma_{R}$ & $\Gamma$ \\
\hline \hline 
$1^{--}$ & $ \rho \pi         $ & P & 100
&  20 &$2^{-+}$ & $ K^{\ast} K       $ & P & 15   &   2  &
$1^{+-}$ & $ \rho \pi             $ & S & 200  &  40\\ \cline{6-10}
	 & $ \omega \eta      $ & P &  30  &   7 &$1^{-+}$
& $ \eta^{'} \eta    $ & P & 30   &$\ast$&
         & $ \omega \eta          $ & S & 100  &  20\\
	 & $ \omega \eta^{'}  $ & P &  15  &   3
&	   & $ K^{\ast} K       $ & P & 15   &   2  &
         & $ \omega \eta^{'}      $ & S & 150  & 30\\ \cline{6-10}
	 & $ K^{\ast} K       $ & P & 30   &   4
&$1^{++}$ & $ K^{\ast} K       $ & S & 125  &  15&
         & $ K^{\ast} K           $ & S & 200  &  30\\ \cline{1-10}
$2^{+-}$ & $ \rho \pi         $ & D &  5   &   1
&$0^{-+}$ & $ K^{\ast} K       $ & P & 60   &   8&
         &                          &   &      &    \\
\hline 
\end{tabular}
\etablec

\btablec
\caption{As in table \protect\ref{flux19302}
but for initial hybrid $s\bar{s}$, and
$\eta (\eta^{'})$ indicating $s\bar{s}$ at 550 MeV
(960 MeV) respectively . }
\label{flux19304}
\begin{tabular}{|l|l|c|r|r||l|l|c|r|r||l|l|c|r|r|}
\hline 
$A$ & $B,C$ & $L$ & $\Gamma_{R}$ & $\Gamma$ &
$A$ & $B,C$ & $L$ & $\Gamma_{R}$ & $\Gamma$ &
$A$ & $B,C$ & $L$ & $\Gamma_{R}$ & $\Gamma$ \\
\hline \hline 
$1^{--}$ & $ K^{\ast} K       $ & P & 90   &  15 &$1^{+-}$
& $ K^{\ast} K           $ & S & 150  &  20&
$1^{-+}$  & $ \eta^{'} \eta        $ & P &  70  &$\ast$\\
	 & $ \phi   \eta      $ & P & 60   &   8 &
& $ \phi   \eta          $ & S & 350  &  40&
          & $ K^{\ast} K           $ & P &  50  &   6\\ \cline{11-15}
	 & $ \phi   \eta^{'}  $ & P & 15   &   2 &         &
$ \phi \eta^{'}        $ & S & 350  &  40&
$1^{++}$ & $ K^{\ast} K       $ & S &  80  &  10 \\ \cline{1-15}
$2^{+-}$ & $ K^{\ast} K       $ & D &  6   &  1 &$0^{-+}$ & $K^{\ast}
K            $ & P & 175  &  30 &
$2^{-+}$  & $K^{\ast} K            $ & P &  40  &   6\\
\hline 
\end{tabular}
\etablec

\end{document}